%% file: LBASS.tex
\documentclass[fleqn,usenatbib]{mnras}

\usepackage[T1]{fontenc}

\DeclareRobustCommand{\VAN}[3]{#2}
\let\VANthebibliography\thebibliography
\def\thebibliography{\DeclareRobustCommand{\VAN}[3]{##3}\VANthebibliography}

\usepackage{graphicx}	% Including figure files
\usepackage{amsmath}	% Advanced maths commands
\usepackage{amssymb}	% Extra maths symbols

\title[L-BASS]{L-BASS: a project to produce an absolutely calibrated 1.4 GHz sky map ~~ I - Scientific rationale and system overview}

\author[D. P. Zerafa  et al.]{
D. P. Zerafa,$^{1}$
P. N. Wilkinson,$^{1}$
C. J. Radcliffe,$^{2}$
J. P. Leahy,$^{1}$
I. W. A. Browne$^{1}$
P. J. Black,$^{1}$\thanks{E-mail: phillip.black@postgrad.manchester.ac.uk}
\\
$^{1}$Jodrell Bank Centre for Astrophysics, %$Alan Turing Building, 
Department of Physics and Astronomy, The University of Manchester, Oxford Road, Manchester, M13 9PL, UK\\
$^{2}$Phase2 Microwave Ltd, Unit 1a, Boulton Rd, Pin Green Ind. Est., Stevenage, SG1 4QX, UK\\
}

\date{Accepted XXX. Received YYY; in original form ZZZ}

\pubyear{2025}

\usepackage{newtxtext,newtxmath}
\begin{document}
\label{firstpage}
\pagerange{\pageref{firstpage}--\pageref{lastpage}}
\maketitle

\begin{abstract}
L-BASS is an instrument designed to produce an absolutely calibrated map of the sky at a wavelength of 21\,cm (L-band) with a radiometric accuracy of $\le 0.1$\,K and with an angular resolution of $23\degr$. The prime motivations are to improve the temperature calibration of higher resolution maps and to investigate the steep spectrum radio background proposed by the ARCADE~2 team. The instrument consists of a pair of conical horn antennas which can scan independently in elevation; each antenna produces a circularly polarized output. The difference in signals from the antennas is measured with a continuous-comparison receiver connected to a digital spectrometer sampling the signal from 1400\,MHz to 1425\,MHz within the protected radio astronomy band. %At any one time one horn will point at the North Celestial Pole (NCP). The first phase of the experiment will produce a map referenced with respect to the NCP. In the second phase we will replace the scanning horn with a cryogenic reference load in order to calibrate the absolute brightness temperature of the NCP.  
 We describe the astrophysical motivation for
the project, the design requirements and how these will be attained.

\end{abstract}


%%%%%%%%%%%%%%%%%%%%%%%%%%%%%%%%%%%%%%%%%%%%%%%%%%

%%%%%%%%%%%%%%%%% BODY OF PAPER %%%%%%%%%%%%%%%%%%

\section{Introduction}

\subsection{Astrophysical Context} 

The discovery of the Cosmic Microwave Background (CMB) by \citet{Penzias1965} highlighted the value of absolute brightness temperature measurements in radio astronomy. At the very start, we should clarify what we mean by `absolute' measurements. The raw output of a telescope can often be represented as
\begin{equation}
   V(t) = G(t) (B*T_{\rm sky} + N(t) + M),
   \label{eq:simple_gain}
\end{equation}
where $G$ is the gain (in output units K$^{-1}$), $B$ is the beam pattern (normalised to unit area), the convolution $B*T_{\rm sky}$ is evaluated in the pointing direction at time $t$, $M$ is a constant offset, and $N(t)$ is the (zero-mean) random noise \cite[e.g.,][]{Planck2015_V}. 

By an absolute measurement we mean that both $M$ and $G$ are determined to useful accuracy\footnote{%
`Useful' depends on the scientific context: for Penzias and Wilson, $\pm 1$\,K in $M$ and better than $\pm 0.1$ in $G$ were sufficient.}, ideally through relatively short chains of comparisons ending in well-defined standards. Evidently, uncertainty in $M$ translates directly into uncertainty in the zero-level component of the sky temperature. Absolute measurements in this sense are rare in radio astronomy: the prime example are measurements of the CMB temperature, pioneered by \citet{Penzias1965} and culminating in the FIRAS experiment on the {\it COBE} space misson \citep{Mather1999,Fixsen1999}; also very relevant to this paper is the ARCADE-2 balloon experiment \citep{Fixsen2011}. In contrast, the DMR on {\it COBE} \citep{Bennett1996}, and {\it WMAP\/}
\citep{Bennett2013}, as differential experiments, do not measure $M$ directly, although \citet{Bennett1996} derive a value ($\pm 20$\,mK) from the {\rm orbital} dipole of the CMB -- about 3 orders of magnitude less accurate than their pixel-level differential measurements. In contrast, {\it WMAP\/} and {\it Planck\/} reverse this process by calibrating $G$ via the orbital dipole to the CMB, as measured by FIRAS -- a process \citet{Planck2018_I} describe as absolute calibration. Although, unlike {\it COBE} DMR and {\it WMAP}, {\it Planck\/} did not record the difference between simultaneously-measured sky positions, it is still essentially a differential experiment, relying on rapid scanning and destriping map-making to give sky maps with arbitrary zero-levels.
Fortunately, a great deal of cosmological information is encoded in the CMB {\em fluctuations}, which is why such data can be used both to test the $\Lambda$CDM model of the early universe and to allow its refinement \citep[e.g.,][]{Planck2018_VI,Planck2018_VII,Planck2018_IX}.

However, the analysis of the spacecraft maps, made at frequencies above 20 GHz, depends significantly on lower frequency maps for accurate removal of the diffuse `foreground' emission from the Milky Way due to synchrotron, free-free and spinning dust emission. The most reliable approach is direct astrophysical modelling of the foreground spectrum of each component, and its variation across the sky \citep[e.g.,][]{Planck2015_X}, which relies on absolute measurements in our sense at lower frequencies, even though the zero-level component of the foreground contamination is irrelevant in the CMB-dominated channels. This has proved difficult \citep[e.g.,][]{Wehus2017} because, like {\it WMAP\/} and {\it Planck}, the ground-based radio telescopes used at lower frequencies also use differential techniques that do not record $M$.  Even when determined in principle, it is often significantly more uncertain than the relative fluctuations. Moreover, ground-based maps, made with dish-type radio telescopes, often contain significant additive artefacts due to ground pick-up via sidelobes, and sometimes also suffer from declination-dependent gain errors. All-sky maps  are \citep[e.g.,][]{Haslam1982,Reich2004} or will be \citep{Jones2018,Wolleben2021} assembled from data taken with different telescopes in the two geographic hemispheres. 
 
Correcting subtle  large-scale errors in current large-area 1.4 GHz maps \citep[e.g.][]{Reich2004,Calabretta2014,Wolleben2021} is one of the prime motivations for us to make an absolutely calibrated L-Band measurement made with horn antennas. In this paper we describe the project to do this, L-BASS, which stands for L-Band All-Sky Survey.

A further motivation for L-BASS arose from the identification by the ARCADE team \citep{Fixsen2011,Seiffert2011} of a new isotropic radio background asserted to be distinct from the CMB, the diffuse emission from the Milky Way and the integrated effect of discrete extragalactic sources. For the identification the ARCADE team added data from published lower frequency maps to their measurements from the ARCADE-2 balloon-borne instrument.  The steep spectrum thus derived is very similar both to that of the diffuse galactic synchrotron continuum and the integrated contribution of extragalactic radio sources. Support for the existence of a background with this spectral index comes from \cite{Dowell} who report measurements made with the Long Wavelength Array in the frequency range 40 to 80~MHz. An explanation in terms the integrated contribution of discrete extragalactic sources can effectively be discounted unless there is a massive new population of faint radio sources below current detection limits. Known radio sources make only a $\approx 20$ per cent contribution to the isotropic background \citep{Vernstrom2011,Vernstrom2015,2023MNRAS.520.2668H}. How much impact the diffuse Galactic emission has on the claimed isotropic background is, however, less clear-cut. 

The evidence for and against a new isotropic background has been discussed in detail during a workshop \citep{Singal2018} at the end of which Singal {\it et al.} summarised three current possibilities as follows:
\begin{itemize}
\item{Zero level offsets in existing maps and the limited sky coverage of ARCADE-2 mean the original interpretation is in error; there is no background other than that provided by the integrated emission from extragalactic radio sources. }

\item{ARCADE and others are correct in isolating a background but the signal is primarily Galactic; this  would challenge the picture of our Galaxy as a typical spiral. }

\item{ARCADE and others are correct in isolating a background and the excess is extragalactic; this would be the most interesting photon field in sky since it is unaccounted for.}
\end{itemize}
The discussion in a recent follow-up workshop \citep{Singal2023} shows that the above summary remains valid.
Emphasising the last of these points, many authors \citep[e.g.][]{Feng2018,Fialkov2019}
have pointed out that an additional population of photons could help to account for the unexpectedly strong absorption feature at $\approx 78$\,MHz identified with atomic hydrogen at redshift $\approx 17$ detected by \citet{Bowman2018}. We note, however, that the reality of this feature has now been disputed by \citet{Singh2022}, and the enhancement mechanism may be ineffective \citep{Acharya2023}. Nevertheless confirmation of a  new isotropic background would carry important implications about the physical conditions in the early universe. \\

\subsection{Observational context}

There is a clear need for large-area absolutely calibrated maps of the sky at a range of frequencies from 100s of MHz to $\sim  5\,$GHz to correct existing maps, to provide new insights into the ISM and to address the reality of a new isotropic radio background. 

Making such absolutely calibrated maps is the stated goal of the GEM project \citep[e.g.][]{Tello2013}. The GEM observations are made with a small 5.5-m dish; although its size facilitates the construction of shielding to mitigate stray radiation, the 2.3-GHz maps reported by Tello et al. illustrate the challenges in removing ground contamination from dish-based measurements. Destriping algorithms were needed and in order to calibrate the GEM map absolutely a cross comparison had to be made with the Rhodes/HartRAO 2365-MHz map (made with a 25-m dish, \citealt{Jonas1998}) whose scale was itself referred back to the absolute measurements at 2\,GHz made by \citet{Bersanelli1994}. The northern C-BASS 5-GHz polarisation survey \citep{Jones2018}, made with a 6.1-m dish whose optics were also carefully designed to mitigate spurious pick up, encountered comparable ground contamination problems to those described by \citet{Tello2013}. This evidence strongly suggests that surveys with dish telescopes, while providing valuable information about the diffuse galactic emission at degree-scale resolution, are unlikely to provide the best absolute radiometric temperature measurements of the sky, except below about 300\,MHz where the sky temperature significantly exceeds the anticipated ground contamination.

Horn antennas, pioneered in this context by \citet{Penzias1965}, can be designed to have ultra-low sidelobe levels and are best suited to the task although practicality limits them to lower angular resolution than dishes. 
 The TRIS team \citep{Zannoni2008,GervasiTRIS,Tartari2008} have made horn-based absolute measurements at 0.6\,GHz, 0.8\,GHz, and 2.3\,GHz with an  $18^{\circ} \times 23^{\circ}$ beam, but only in a single declination strip and only the 0.6-GHz data have a level of uncertainty ($\pm$0.15\,K) appropriate to make a contribution to meeting our astrophysical goals.

\subsection{The general approach}

Our ultimate aim is to map the entire sky at 1.4\,GHz with $\approx 23^\circ$ resolution and an absolute radiometric accuracy of $\le 0.1$\,K. This target accuracy is driven by the fact that at 1.4\,GHz the excess background emission identified by the ARCADE team is predicted to be $\approx0.5$\,K.  For the reasons outlined above L-BASS uses horn antennas which  should have as low sidelobes as possible and have well-defined circular beams. Since the sky emission is partially linearly polarized it is desirable to work with circular polarization. The polarizer outputs are connected to a  continuous comparison (sometimes called pseudo-correlation) receiver whose architecture is similar to that used by {\it WMAP\/} \citep{Bennett2003} and {\it Planck}-LFI \citep{Mandolesi2010} and its output gives the difference between the signals from the two horns. One of the horns will be pointed to the North Celestial Pole (NCP) which acts as the sky radiometric reference temperature.
 
In order to achieve absolute calibration the radiometric temperature of the NCP will then be measured by replacing one of the horns with a reference load cryogenically cooled to a physical temperature $\sim 10$\,K -- as close as possible to the predicted radiometric temperature of the NCP (including atmospheric emission) which is $\approx 6.2$\,K.  

The challenge of absolute measurements at the required level is to reduce systematic effects rather than to minimise thermal noise. The active receiver components therefore do not need to be cryogenically cooled, which reduces the project cost. For the same reason observations can be made using a relatively small bandwidth chosen to avoid RFI.  Even in the absence of RFI, great care is needed to measure and/or reduce the “local” radiation foregrounds arising in i) the atmosphere; ii) the surrounding ground and trees; iii) the passive components ahead of the radiometer itself.  In respect of i), observations will only be made in clear sky conditions when the Sun is below the horizon; the atmospheric absorption is well-understood at 1.4\,GHz and the precipitable water vapour content along the beam directions will be monitored with infra-red detectors (see Section \ref{sec:atmosphere_monitor}) In respect of ii),  the instrument is surrounded by a large ground screen which greatly reduces stray radiation.  
In respect of iii), the passive losses of components in front of the first LNAs have been carefully measured in the laboratory prior to deployment and the spatial and temporal variations in the physical temperatures in the passive components will be continuously monitored.  These data allow the losses in and concomitant thermal emission from the passive components to be subtracted from the sky signals.   

In what follows, we provide a broad overview of the experimental aspects of the L-BASS project. In Section~\ref{sec:instrument} we give an outline description of the instrument as a whole.  In Section~\ref{sec:calibration} we discuss the overall calibration strategy and in Section~\ref{sec:observations} describe the different observing modes. Finally, in Section~\ref{sec:conclusions}, we  summarize the current state of the project.\\

\section{The L-BASS instrument}
\label{sec:instrument}

In this section we describe the L-BASS instrument as built at the Jodrell Bank Observatory and discuss how we intend meet the design challenges outlined above. The detailed design of the receiver and the characterization of its constituent components will be presented in Paper II (Zerafa et al. in prep). 

\begin{figure*}
    \centering
    \includegraphics[width=\textwidth]{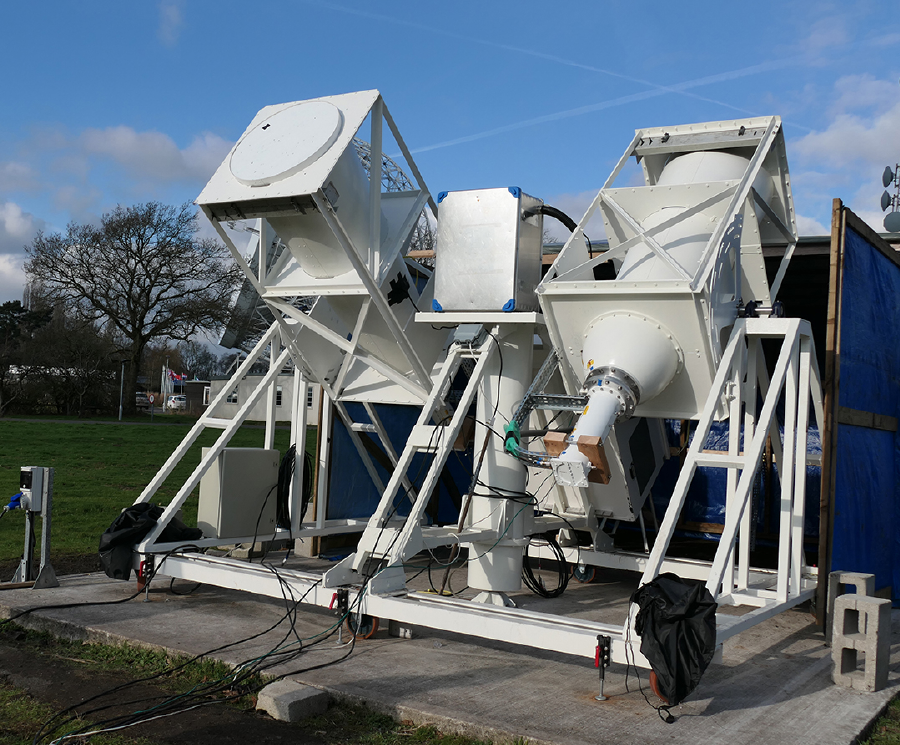}
    \caption{The L-BASS telescope at Jodrell Bank Observatory before erection of the ground screen. The plastic foam endcaps of the horns will not be used in observations, as discussed in Sec.~\ref{sec:weatherproofing}}
    \label{fig:telescope}
\end{figure*}

In Figure~\ref{fig:telescope} we show a picture of L-BASS  taken before the construction of the ground screen. The telescope consists of two identical horns whose  mouth diameters are 0.78\,m and whose lengths, including the circular polarizers at their base, are 3.36\,m.  The resulting FWHM of the beam is $\approx 23\degr$ at 1.4\,GHz. The horns are mounted between two outer A-frames and a fixed central pillar.  During the initial set-up and testing period the entire instrument could be rotated around the central pillar. For mapping purposes the instrument will observe at meridian transit with each horn able to be moved independently in elevation. The horns are connected by flexible RF cables to the receiver which is mounted in a temperature controlled box fixed to the support structure. These 4.04-m cables are clad in commercial pipe insulation and wrapped with heating wire in order to linearize the temperature profile between the heated receiver box at $\sim 35 \degr$C, the cables exposed to the night-time ambient air, and the warmed polarizers at $\sim 15 \degr$C (see Sec.~\ref{sec:weatherproofing}).   

\begin{figure*}
    \centering
    \includegraphics[width=\textwidth]{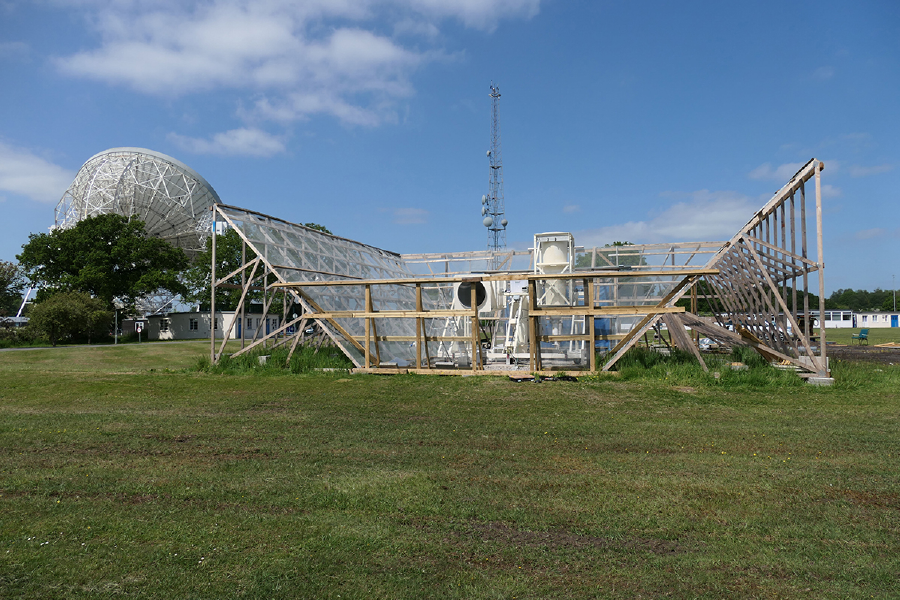}
    \caption{The L-BASS telescope surrounded by its ground screen. Note that the top of the zenith-pointing horn is $\sim 0.4$ m below the top of the ground screen.}
    \label{screen}
\end{figure*}

In order to minimize the instrument's exposure to RFI, all observations will be made in the radio astronomy protected band 1400 to 1427\,MHz. Both to further reduce RFI and to minimize ground pickup a ground screen has been constructed; the west, north and east sides are 4\,m high but to the south it is only 2\,m in order to enable low declination observations. The wooden structure of the screen shown in Figure~\ref{screen} is covered with wire mesh with 6\,mm square holes. 

Key to the success of the instrument are the horns. A modified Potter design  gives low levels of sidelobes, but over a narrow band  \citep{Leech2011} which is not a problem for L-BASS. A Potter horn has a circular cross section with a single step in diameter but we have adopted a design evolved from that of Leech {\it et al.}, but with eight steps rather than three. Simulations indicate that sidelobes should be at $\sim-60$\,dB level. 
Measurements with a feed-test facility and on the sun show that the beam FWHM is 23 degrees, and the sidelobes away from the main beam are at least -40 dB below the peak.
More details of the horn design and testing are presented in Paper II.

Instruments designed to measure/map the absolute brightness temperature of the sky must compare the signal from an antenna with the signal from cold reference load. This can be done using a single antenna and reference load or, as in the approach we have adopted one can divide the mapping and the absolute calibration into separate tasks. In Phase 1, one horn will scan the sky while the other points at the NCP, whose emission gives a fixed radiometric reference signal, provided we correct for the atmosphere (Sec.~\ref{sec:atmosphere_monitor}). 

The pole can be observed continuously without moving an antenna and, as mentioned previously by observing in circular polarisation we ensure that the emission from the NCP will be constant despite the sky being partially linearly polarised and rotating in parallactic angle. The L-BASS polarizers provide two ports that offer circularly polarized outputs, one of which will be connected to the receiver, the other we have repurposed as an input for a Continuous Wave (CW) injection system for gain calibration (Sec. \ref{running gain cal}). These polarizers have been extensively tested in the lab and meet the desired specification. Again, full details will be presented in Paper II.

After Phase 1 is completed,  
in Phase 2 we will measure the radiometric  temperature of the NCP by comparing its emission with that from a cryogenically cooled reference load. There are practical reasons for adopting this two stage approach.  First we wanted to avoid the cost and complication of running a cooled system for the whole $\sim$ 2-year period over which observations will be taken. Our strategy is to take mapping data only on cloudless nights ($\sim 20 \%$ in the UK and entirely unpredictable) thus the system needs to be ready to observe  throughout the year on a night-by-night basis. Secondly a map made relative to the NCP without absolute calibration can itself be of astrophysical interest and could be accurate at the millikelvin level if it were thermal noise limited, rather than systematic error limited.  Even with our best efforts it is certain that the absolutely calibrated map we produce will be limited by systematic errors significantly above the thermal noise level. 

\begin{figure}
    \centering
    \includegraphics[width=8cm]{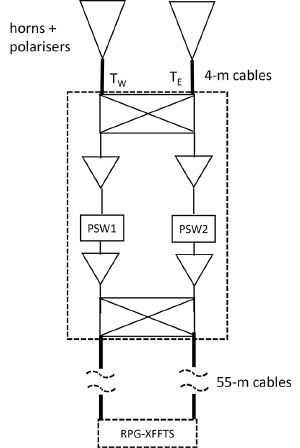}
    \caption{A simplified schematic of the L-BASS system. The two horns are connected to a hybrid by 4-m cables. After that there are amplifiers in both arms. There are two phase switches (PSW1 and PSW2) one in each arm, one of which is driven and the other included just to balance the arms. Following the second hybrid the signals are transferred to the digital spectrometer by 55-m cables.}
    \label{system}
\end{figure}

As described in more detail in Paper II, the receiver is housed in a temperature-controlled 
box mounted on the telescope structure. A schematic diagram of the  layout of the whole system is shown in 
Figure~\ref{system}. The receiver outputs are connected by a pair of $\approx 55$\,m cables to a digital spectrometer manufactured by RPG\footnote{eXtended Fast Fourier Transform Spectrometer (XFFTS) manufactured by Radiometer Physics GmbH (RPG).} 
situated in a nearby building. We use a continuous-comparison receiver because they are designed to suppress the effects of amplifier $1/f$ noise, the random gain fluctuations that arise in all amplifiers. This is achieved  by using a magic-tee hybrid (hereafter 'hybrid') to combine the two incoming signals so that both pass through all the amplifiers before they are recombined by the second hybrid to produce two outputs proportional to the original two inputs. Because both second hybrid output signals will have passed through all the amplifiers they will, in a perfectly balanced system, suffer from the same $1/f$ noise so that when differenced the $1/f$ noise will cancel out. However, this cancellation only works for the effects of components between the two hybrids.  
Any gain changes/losses after the second hybrid, or within the spectrometer, will affect the recorded signals differently. For this reason the standard approach is to include a 180\degr\ ($\pi$) phase switch in between the hybrids, the effect is to swap back and forth the signals in the two outputs of the second hybrid. In L-BASS the switching is done at 10~Hz which is sufficiently fast that combination of  losses in the 55-m cables and gains in the spectrometer do not change significantly within a switch period.  The result is that the receiver produces four data streams corresponding to the 0 and $\pi$ phase switch states of each output channel. 

Because the receiver is not cooled and there is significant attenuation in the 4-m cables ahead of the receiver inputs, the on-sky system temperature, $T_{\rm sys}$ is $\approx 135$\,K, giving a thermal noise level of $\approx 5$~mK for one minute's integration time. This is ample for our purposes, given the large beam: one hemisphere comprises about 36 antenna beam areas \footnote{The antenna beam area is given by $\Omega_A = 4\pi/G$ where $G$ is the antenna gain. EM modeling (see Paper II), gives $G$ = 18.6 dB corresponding to 72.4 beams over $4\pi$ steradians and thus 36.2 beams per hemisphere.}, so we would reach our goal sensitivity of 1\,mK beam$^{-1}$ in $\sim 15$\,hr. In practice we expect to observe for much longer than this to beat down residual systematic errors.

 The spectrometer samples both outputs from the receiver and provides 55\,kHz-wide frequency channels. As in the {\it WMAP\/} and {\it Planck}-LFI systems these two powers are double-differenced \citep{Seiffert2002,Jarosik_2003} to produce a signal that is proportional to the difference between the brightness temperature of the NCP and that point in the sky sampled by the second horn. As noted above while a perfectly balanced continuous-comparison receiver eliminates the effects of gain variations between the hybrids, double differencing can eliminate the effects of gain variations after the second hybrid either in the $\approx 55$\,m connection cables or in the digital spectrometer itself.  During the commissioning observations drift scans will made at constant elevation  but in normal operation one or other of the horns will be continuously scanned in elevation.

The design of the telescope is symmetric so that at any time one of the horns can be fixed to point at the NCP while the other scans in elevation. To check for systematics we will swap the roles of the horns from time to time.

\subsection{Weatherproofing}
\label{sec:weatherproofing}

When not in use, the L-BASS instrument is protected by a wooden, tarpaulin-covered structure, the `hutch', that can be rolled over it and closed at the front with double doors -- this is visible behind the instrument in Fig.~\ref{fig:telescope}. A northern section of the ground screen can be lowered, drawbridge-style, allowing the hutch to be pushed away during observations. Our original plan was to seal the horns with microwave-transparent foam plugs and actively pump dry air through the interior of the horns and polarizers to keep out moisture. However, dew formation on the foam plugs produced unacceptable thermal emission and the plugs and dry-air system were replaced by a heating system: the horns and polarizers are actively heated  to well above the dew point, using a combination of small commercial fan heaters and heating wire, to prevent dew formation. This also reduces temperature gradients along the cables, which connect the polarizer to the warm receiver box, thereby improving the accuracy of our detailed temperature modelling (Sect.~\ref{sec: system noise calibration}). Losses associated with the horn and throat section of the antenna were below the threshold of detectability during lab testing (see Paper II in prep). Therefore we do not anticipate significant thermal emission caused by heating of the horns. We will report the results of differential heating tests of the horns in Paper III (in prep). As discussed in Section~\ref{sec:atmosphere_monitor}, the instrument will not be used when rain or even significant cloud cover is predicted, and so will be protected by the hutch.

\section{Calibration strategy}
\label{sec:calibration}

In this section we give an overview of our approach to the calibration of the L-BASS system. In Paper II we describe the detailed design and performance verification of the different sub-systems and in Paper III (Black et al. in prep) we give a full account of the performance of the interrelated calibration schemes and verify that L-BASS can meet its science goals.   

\begin{figure*}
    \centering
    \includegraphics[width=\textwidth]{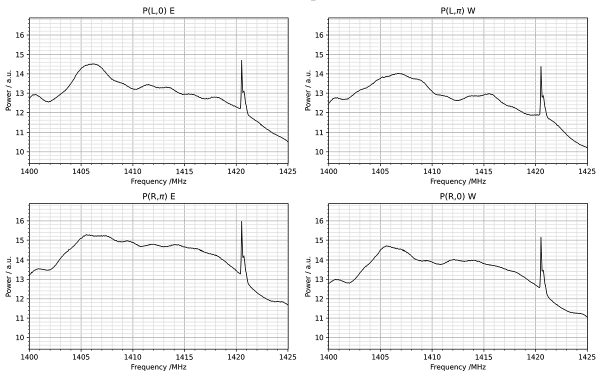}    
    \caption{Example of the uncalibrated bandpasses. Right (R) and Left (L) refer to different inputs to the spectrometer. The two phase switch states are distinguished by $\pi$ and 0. Depending on the phase switch state signals are received from the East (E) and the West (W) horn. These data were taken with both horns pointing at the Zenith, not far from the Galactic plane, giving a particularly strong 21-cm Galactic hydrogen signal, but the line is detectable in all directions.}
    \label{bandpass}
\end{figure*}

The spectrometer arrangement produces four outputs ($i=1$ to 4) corresponding to two phase states (0 and $\pi$) for each output of the second hybrid (and corresponding spectrometer input), labelled left (L) and right (R): example (uncalibrated) spectra from the four outputs are shown in Fig.~\ref{bandpass}. Galactic 21~cm hydrogen emission is clearly detected in these spectra. The ultimate goal of calibration is to enable the four outputs to be translated into a measurement of
the difference in brightness temperature of a patch of sky compared with that from a reference signal which in L-BASS Phase 1 is the emission from the NCP. Two difference-signals are available, one for each phase of the phase switch and the final result we require the average of those two differences.
In the following it is useful to think of different calibration tasks; one is to establish a 
conversion factor of the arbitrary output units produced by the spectrometer into Kelvin, Another is to keep track of the
running gain and also the system noise, both of which will change with time. For practical reasons establishing the conversion factor is a two-stage process. Initially a working value needs to be established until in Phase 2 of the project when an absolute value is obtained by comparison of the NCP signal with that from a cryogenically cooled reference load. The initial working value is required so that corrections to the system noise can be implemented using a ``passives model'' -- see below.

\subsection{Establishing the preliminary brightness temperature scale}

There are two astronomical sources of emission that are easily detected by L-BASS--the Sun and Galactic H1 emission. For the Sun there are daily flux density measurements at 1415 MHz made at the Learmonth Solar Observatory \citep{Learmonth} whose absolute accuracy is around 5\%; these provide us with $S_{\rm sun}$. Comparared with the L-BASS beam the Sun is effectively a point source\footnote{\cite{SOLAR} give a formula for the effective circular diameter of the Sun at $\nu$ GHz = 32.0 + $\nu^{-0.6}$ arcmin. At 1.415 GHz this corresponds to 33.8 arcmin and thus a solid angle = $7.59 \times 10^{-5}$ sterad; this is to be compared with the solid angle of the antenna beam $\Omega_{\rm A}$= 0.173 sterad.} and hence its brightness temperature $T_{\rm sun} = S_{\rm sun}A_{\rm eff}/2k$ where $A_{\rm eff}=\lambda^{2}/\Omega_{\rm A}$.  
During Phase 1  $T_{\rm sun}$ is typically 100K-150K and hence pointing at the Sun
roughly doubles the system temperature seen on cold sky. The receiving system is linear over this range (see Paper II).

An independent approach is to compare the HI signals detected by L-BASS with the expected results from existing well-calibrated HI maps (e.g. HI4PI \citep{HI4PI} and LAB \citep{LAB}) after correcting them to match both the angular resolution and frequency resolution of L-BASS. When convolved with a 23 degree beam, the brightness temperature of HI emissions away from the Galactic centre are  < 50~K (above the continuum power). 
\\
In phase 2 the absolute zero-level and scaling factor will be established using a cryogenic comparison load.  The temperature of the load can be commanded to change in an approximate range of 15<$T$<40~K for the purposes of calibration.  This will allow us to verify the system's linearity of response to sky source temperatures.  
    
\subsection{Running gain calibration}
\label{running gain cal}
The running gain is
complicated because just about everything from polarizers, 4-m cables, receiver
(including phase switch), 55-m cables and the spectrometer all make their contribution\footnote{Any attenuation in front of the first low noise amplifiers adds to the system noise as well as reducing the signal strength. This added noise will be discussed in the next section.} . As
stated above there is not one running gain but four slightly different ones as the two
horn signals in the two different phase states will have subtly 
different gains. The plan is to inject CWs into the two horns. The CW frequency will be different for each horn; thus knowing the strengths of the injected CWs the actual strengths recorded by the spectrometer 
will depend only on the cumulative effect of all the above gain terms. By splitting the CWs before injection and  feeding  half the signal to power meters the strengths of the CWs can be monitored continuously. High powers are  required to drive the  power meters but only very low powers should reach the receiver. Conveniently each polarizer has an unused port and a highly attenuated  CW signal can be injected into the unused ports of each polarizerand hence follow the same post-polarizer path as the sky signal through the receiver. Each of these leaks about -30 dB into the other output of the polarizer. The intention is to make use of these otherwise redundant ports but  before injection into the receiver a further -50 dB of attenuation is required so as not to saturate the receiver. 

 \subsection{System noise calibration}
 \label{sec: system noise calibration}
As with gains there are four slightly
different noises that we need to take into account. The dominant contributors to the
system noises are the first LNAs and everything skywards of them, particularly the 4-m cables. All these
contributions can change with time but by far the largest changes will come from the
temperature changing in the 4-m cables which are not in temperature-controlled environment. Because we need to know this temperature-dependence of the cable attenuation we have made laboratory measurements of this property. Temperature sensors are fixed at 7 places along each cable as well in the receiver box and along the length of the polarizers. Knowing the attenuation temperature coefficients of the cables, and of the other components, these can be built into a model (the passives model) to enable the added noise to be predicted. As the predicted value will be in Kelvin any calibration correction to the receiver noise needs to be after the conversion factor to Kelvin has been applied to the spectrometer output.  

Another potential approach to system noise calibration is to use frequent measurements of the strength of the detected signal when pointing at the NCP. If the sky signal
is constant, and a gain correction has been made using the CWs, any residual
changes in recorded spectrometer output must result from changes in the cumulative
factors that contribute to the system noise. With a constant sky signal it becomes possible 
to track these system noise changes. In practice the detected signal from the NCP will vary slightly because of atmospheric attenuation and a small correction will be required taking into account the atmospheric temperature, pressure and humidity   as measured on site.  The challenge to make this method of calibration to work well is to be able return the scanning horn to observe the NCP on a  short enough timescale that the system noise has not had a chance to change significantly during the process. This will be possible only when motor-driven scanning is implemented. Before that is implemented system noise calibration will rely on results from the  passives model.

\label{sec:gain_cal}

\section{Observing modes}
\label{sec:observations}
\subsection{Mapping observations}

Initial observations have started using meridian drift scans at constant elevation but the intention is to move to a nodding observing mode with one antenna moving from looking at the NCP to a lower elevation limit and back. With the symmetrical design of the system, observations can be done with either of the horns being used as a reference and repeating observations with different reference horns will reduce systematic errors. At the current location at Jodrell Bank the lower elevation limit for useful observations is set to  $\sim +40\degr$ by tall trees to the south of the telescope. This corresponds to declination $\sim +10 \degr$ for the centre 
of the beam. 
The plan is that, once maps of most of the northern sky have been made from Jodrell Bank, the instrument will be moved to a site further south, say at latitude $\sim 30\degr$, and with a clear southern horizon so that it will be possible to map 75 per cent of the sky. A repeat of the area mapped from Jodrell will provide a useful consistency check on our results.

The Sun is the strongest source in the sky, which almost doubles the system noise when pointed at directly. 
To avoid contamination by solar emission our primary mapping observations will be made after sunset and before sunrise\footnote{An added advantage of observing at night is the absence of direct sunlight which can produce differential heating of horns and cables.}.  However, as stated above solar transit observations are one of the ways to calibrate the system noise power in order to put our observations on a radiometric temperature scale.  

Emission from the Galactic 21-cm hydrogen line will be present in all our observations. For continuum mapping the spectrometer channels containing hydrogen emission will be removed. However, the line data will be kept 

for calibration purposes.  Whether our observations employ drift-scanning or motorised nodding, the beam of the scanning horn will periodically cross the Galactic plane.  These crossings will provide high signal-to-noise HI measurements which can be compared to data from the HI4PI survey \citep{HI4PI}, offering a further route to track gain changes in the system.  We anticipate that following the absolute calibration of L-BASS measurements with a cryogenic load, that our work may offer refinements and corrections to the brightness scale and stray light mitigation of HI surveys such as HI4PI.

\input{L-BASS_Infrared}

\input{Where_next}

\section*{Acknowledgements}

DZ acknowledges support from an STFC studentship. Construction of L-BASS was part-funded by STFC grant ST/L000768/1. We thank Ralph Spencer for designing the phase switch driver circuit and Clive Dickinson for his work on the spectrometer software. Without the sterling efforts of the Jodrell Bank technical staff, particularly John Edgley, Adrian Galtress, Dave Clark and Alan Williams, the project would not have been possible. We also thank Jordan Norris for his work to implement the CW calibration system, and Jens Chluba for useful discussions.  PJB acknowledges support from the UoM Department of Physics \& Astronomy via a Department Scholarship.  Finally we would like to thank the anonymous reviewers for kind words and a constructive critical appraisal of this work.  Feedback about the technical description of the instrumentation, flux scale calibration and beam areas aided us in presenting L-BASS with greater clarity and specificity.

\section*{Data availability statement}

This is an announcement paper; as such, no data or data analysis is included in this work other than that already included in the article.  More detail and supporting data will follow in papers II \& III by Zerafa et al. and Black et al. (both in prep).

%%%%%%%%%%%%%%%%%%%%%%%%%%%%%%%%%%%%%%%%%%%%%%%%%%



\bibliographystyle{mnras}
\bibliography{lbassbiblio} % if your bibtex file is called example.bib

%\begin{thebibliography}{99}
%\end{thebibliography}

%%%%%%%%%%%%%%%%%%%%%%%%%%%%%%%%%%%%%%%%%%%%%%%%%%

%%%%%%%%%%%%%%%%% APPENDICES %%%%%%%%%%%%%%%%%%%%%

%%%%%%%%%%%%%%%%%%%%%%%%%%%%%%%%%%%%%%%%%%%%%%%%%%

%\input{AppendixA}
%\input{AppendixB}
% Don't change these lines
\bsp	% typesetting comment
\label{lastpage}
\end{document}

%% file: L-BASS_Infrared.tex
\subsection{Atmospheric Monitoring}
\label{sec:atmosphere_monitor}
To achieve our science goals the absorption and the concomitant thermal emission from the neutral atmosphere must be must be known, the latter to well below 0.1\,K. A fully quantitative accounting of atmospheric effects on L-BASS measurements will be given in Paper III; here we limit ourselves to a broad overview of how we will proceed.

At L-band the dominant contribution to the absorption coefficient comes from well-mixed molecular oxygen; the contribution from poorly-mixed water vapour plus liquid water in clouds is much smaller but more variable and harder to predict. \citet{6276237} aimed their radiative transfer model specifically at the 1400--1427\,MHz band and compared its predictions with data from many radiosonde flights; the averaged atmospheric brightness contribution at the zenith is close to 2 K. This is consistent with the average zenith brightnesses at 2.3\,GHz for the JPL Deep Space Network sites \citep{123456} scaled to sea level and to 1.4\,GHz.  To take account of varying conditions and pointing angles we will use the Recommendation ITU-R~P.676-13 (dated 08/2022) which lays out the equations required to calculate the gaseous attenuation along slant paths; software to implement the equations is publicly available\footnote{https://itu-rpy.readthedocs.io/en/latest/}. 

Surface meteorological measurements for temperature, pressure and humidity are taken by a dedicated weather station co-located with the L-BASS instrument.  These will be used to update the pressure-dependent oxygen and water vapour attenuation values as a function of time.  We also intend to monitor the wet atmospheric conditions for the elevation angle of each antenna using data taken with a pair of Eurotech Aurora Cloud Sensors\footnote{see http://www.auroraeurotech.com/CloudSensor.php} (EACS) designed for the amateur astronomy market. One of the sensors in the EACS is a Melexis MLX90614xAA which covers the band 5.5\,$\mu$m to 14\,$\mu$m. The data from these IR sensors are written as temperatures in CSV files and imported into a spreadsheet for each observing run\footnote{The required modifications to the EACS were made by staff at Qinetic PLC for a different application.}.

Given the low opacity of water vapour it is only necessary to establish the precipitable water vapour (PWV) content to within a few mm to be able to calculate the emission to $<0.01$K. It has been shown \citep{MeasuringTotalColumnWaterVaporbyPointinganInfraredThermometerattheSky} that a clear sky IR temperature, measured with low-cost IR sensors similar to that in the EACS, predicts the PWV to the required accuracy.

Liquid water in clouds poses more of a problem since \citet{Staggs1996} showed that that heavy cloud can contribute $\ge 0.1$\,K at 1.4\,GHz. L-BASS observations will therefore be restricted to largely cloud-free skies. To help decide if observing conditions are suitable, the difference between the surface temperature and the IR sky temperature provides a simple diagnostic of the level of cloud cover along a line of sight. 
In overcast conditions the optically thick cloud base is typically at an altitude $1.0\pm 0.5$\,km thus, with the wet adiabatic lapse rate of $\sim 6.5$\,K\,km$^{-1}$, the cloud base IR temperature is invariably within $10$\,K of the surface temperature. By contrast in clear sky conditions the zenith IR temperature is typically $\ge30$\,K below that of the surface at our observing site. This temperature difference is sufficiently large that the onset of significant cloud cover during unattended observations is easy to recognise.

%% file: where_next.tex
\section{Conclusions and future work.} 
\label{sec:conclusions}

We have built an instrument which will be capable of producing a low angular resolution map of the sky at L-band absolutely calibrated to better than 0.1\,K.  The key features of the instrument and its operation are:

\begin{itemize}

    \item A symmetrical twin horn architecture and a receiving system which measures the difference between the signal coming from a scanning horn and the other pointing at the NCP as a reference. 

    \item Observations on the meridian in the protected radio frequency band between 1400~MHz and 1427~MHz.  
    
    \item Horns with with sidelobes $\leq$--40~dB complemented by a ground screen to minimize stray radiation.
    
     \item An extensive physical temperature monitoring network combined with temperature control of critical components.  

     \item A CW calibration system which continuously measures the overall system gain to high accuracy.

     \item A phase switching constant comparison (pseudo-correlation) receiver to produce the signals to be differenced. The signals are processed in a digital spectrometer enabling RFI to be identified then flagged and Galactic neutral hydrogen and CW calibration signals to be measured. 

     \item Observations restricted to cloudless nights with local meteorological data enabling atmospheric absorption to be modelled with an accuracy $\ll 0.1$ K.   

    \item The second phase of the project is to measure the absolute brightness temperature of the NCP emission during which one of the horns will be replaced by a cryogenically cooled reference load.

\end{itemize}

The instrument has been assembled at the Jodrell Bank Observatory and is in the final stages of commissioning. Once most commissioning has been completed in spring 2025 the first science observations will be scans at fixed elevation 
aimed to cover both the minimum in Galactic emission (RA $\approx 10^{\rm h}$, $\delta \approx 32\degr$) and the declination strip covered by the TRIS team at 600~MHz \citep{Zannoni2008,GervasiTRIS,Tartari2008}. This will also cover part of the region mapped by ARCADE-2. %thus enabling consistency checks to be made. 
Once elevation drives have been implemented the aim is to map the Northern sky above declination $+10\degr$ which is the practical limit of the present site. Finally, the plan is to move the whole system further south to a site  
still in the northern hemisphere but with an unobstructed southern 
horizon. From such a site it should be possible to map the sky down to declination $-20\degr$. In the long term we plan to move to the southern hemisphere to obtain full sky coverage.